\pdfoutput=1

\documentclass[preprint,preprintnumbers,amsmath,amssymb,floatfix,endfloats*]{revtex4}

\usepackage{graphicx}
\usepackage{dcolumn}
\usepackage{bm}
\usepackage{amsfonts}

\DeclareMathAlphabet{\mathsfsl}{OT1}{cmr}{bx}{it}
\begin{document}
\title{Accelerated rejuvenation in metallic glasses subjected to elastostatic compression along alternating directions}
\author{Nikolai V. Priezjev$^{1,2}$}
\affiliation{$^{1}$Department of Mechanical and Materials
Engineering, Wright State University, Dayton, OH 45435}
\affiliation{$^{2}$National Research University Higher School of
Economics, Moscow 101000, Russia}
\date{\today}
\begin{abstract}

The influence of static stress and alternating loading direction on
the potential energy and mechanical properties of amorphous alloys
is investigated using molecular dynamics simulations. The model
glass is represented via a binary mixture which is first slowly
annealed well below the glass transition temperature and then
subjected to elastostatic loading either along a single direction or
along two and three alternating directions. We find that at
sufficiently large values of the static stress, the binary glass
becomes rejuvenated via collective, irreversible rearrangements of
atoms. Upon including additional orientation of the static stress in
the loading protocol, the rejuvenation effect is amplified and the
typical size of clusters of atoms with large nonaffine displacements
increases. As a result of prolonged mechanical loading, the elastic
modulus and the peak value of the stress overshoot during startup
continuous compression become significantly reduced, especially for
loading protocols with alternating stress orientation. These
findings are important for the design of novel processing methods to
improve mechanical properties of metallic glasses.

\vskip 0.5in


\end{abstract}

\maketitle

\section{Introduction}

The development of novel fabrication techniques and processing
methods that suppress crystallization and oxidation of metallic
glasses is important for various biomedical and structural
applications~\cite{Forming20}.  Although metallic glasses are known
to possess exceptionally high yield stress and relatively large
elastic strain limit due to their amorphous structure, upon further
increasing stress, the plastic deformation becomes sharply localized
within narrow regions, called \textit{shear bands}, leading to
material failure~\cite{Greer16}. Notably, it was recently
demonstrated that upon triaxal compression at room temperature,
metallic glasses can be rejuvenated and exhibit
\textit{strain-hardening}, which inhibits the formation of shear
bands and catastrophic failure during uniaxial tension or
compression tests~\cite{Greer20hard}.  In general, common methods to
rejuvenate amorphous alloys and improve their plasticity include
irradiation, cold rolling, high-pressure torsion, flash
annealing~\cite{Ogata15,Maass18,Priez19one,WangTang20}, elastostatic
loading~\cite{Park08,Jae-Chul09,Tong13,Wang15,Bai15,GreerSun16,
Zhang17,PanGreer18,Samavatian19,PriezELAST19}, and more recently
discovered, cryogenic thermal
cycling~\cite{Ketov15,Guo19,Mirdamadi19}.   It was originally shown
that a particularly simple way to induce rejuvenation and improve
mechanical properties of metallic glasses is to apply a static
stress slightly below the macroscopic yield stress~\cite{Park08}. In
particular, the prolonged static loading within the elastic range
induces irreversible rearrangements of atoms in a disordered solid
and relocates the system into a higher potential energy state with a
lower density~\cite{GreerSun16}. More recently, it was found using
atomistic simulations that binary glasses undergo a
relaxation-to-rejuvenation transition upon increasing uniaxial
static stress at about half the glass transition
temperature~\cite{Zhang17,PriezELAST19}.  However, the influence of
alternating loading direction, multiaxial constraint, and
hydrostatic pressure and temperature during elastostatic loading of
amorphous alloys on their energy states and mechanical properties
remains largely unexplored.

\vskip 0.05in

During the last decade, a number of molecular dynamics simulation
studies have been carried out to investigate the atomic structure
and mechanical properties of amorphous materials subjected to
stress- or strain-driven periodic deformation~\cite{Lo10,Priezjev13,
Sastry13,Reichhardt13,Priezjev14, IdoNature15, Priezjev16,
Priezjev16a, Kawasaki16, Sastry17, Priezjev17, Priezjev18,
Priezjev18a, NVP18strload, SastryPRX19, PriMakrho05, PriMakrho09,
NVP19alt, Priez19ba, Jana20, Peng20, Priez20altY, Pelletier20,
Priez20del}.   Interestingly, it was demonstrated that at zero
temperature atomic trajectories in a disordered solid under
low-amplitude oscillatory shear eventually fall into the so-called
`limit cycles' and become exactly repeatable after one or more shear
cycles~\cite{Reichhardt13,IdoNature15}.  In the presence of thermal
fluctuations, rapidly quenched glasses subjected to periodic shear
strain below the yielding point gradually relax towards low energy
states with nearly reversible dynamics, typically during thousands
of cycles, in a process termed `mechanical
annealing'~\cite{Priezjev18,Priezjev18a,NVP19alt,Jana20,Priez20del}.
More recently, it was shown that at low temperatures, the relaxation
process can be accelerated in periodically driven metallic glasses
if the shear orientation is alternated along two or three spatial
directions~\cite{NVP19alt}.   It was later found that when an
additional shear orientation is introduced in the loading protocol,
the number of transient cycles before the yielding transition is
reduced but the critical strain amplitude remains
unchanged~\cite{Priez20altY}. Despite this progress, however, the
optimization of thermomechanical processing methods requires further
investigation of the operating conditions and details of loading
protocols in order to access a wider range of energy states and
improved properties of amorphous alloys.

\vskip 0.05in

In this paper, the influence of the static stress and variable
loading direction on the potential energy and mechanical properties
of disordered solids is investigated using molecular dynamics
simulations. We study the binary mixture slowly annealed below the
glass transition temperature and then loaded statically ether along
a single direction or along two and three spatial directions. It
will be shown that with increasing static stress, the potential
energy increases and the size of plastically deformed regions
becomes larger.    Remarkably, the effect of rejuvenation is
enhanced when an additional orientation of the static stress is
included in the loading protocol. Furthermore, the results of
mechanical tests after elastostatic loading indicate that both the
elastic modulus and yield stress are reduced in rejuvenated samples.

\vskip 0.05in

The rest of the paper is divided into three sections. The details of
molecular dynamics simulations and the loading protocol are
described in the next section. The time dependence of the potential
energy and mechanical properties as well as the analysis of
nonaffine displacements are presented in section\,\ref{sec:Results}.
The main results are briefly summarized in the last section.

\section{Molecular dynamics (MD) simulations}
\label{sec:MD_Model}

In our study, the amorphous alloy is represented via the
Kob-Andersen (KA) binary mixture of two types of atoms with strong
non-additive interaction that suppresses crystallization upon
cooling below the glass transition
temperature~\cite{KobAnd95,Dyre19}. The parametrization of the KA
model is similar to the one used by Weber and Stillinger to study
the amorphous metal alloy
$\text{Ni}_{80}\text{P}_{20}$~\cite{Weber85}. In the KA model, the
atoms of types $\alpha,\beta=A,B$ interact via the Lennard-Jones
(LJ) potential as follows:
\begin{equation}
V_{\alpha\beta}(r)=4\,\varepsilon_{\alpha\beta}\,\Big[\Big(\frac{\sigma_{\alpha\beta}}{r}\Big)^{12}\!-
\Big(\frac{\sigma_{\alpha\beta}}{r}\Big)^{6}\,\Big],
\label{Eq:LJ_KA}
\end{equation}
with the standard parametrization for the energy and length scales:
$\varepsilon_{AA}=1.0$, $\varepsilon_{AB}=1.5$,
$\varepsilon_{BB}=0.5$, $\sigma_{AA}=1.0$, $\sigma_{AB}=0.8$,
$\sigma_{BB}=0.88$, and $m_{A}=m_{B}$~\cite{KobAnd95}. To speed up
the computation, the cutoff radius of the LJ potential is set to
$r_{c,\,\alpha\beta}=2.5\,\sigma_{\alpha\beta}$. The system consists
of $N=60\,000$ atoms. Throughout the study, the physical quantities
are reported using the LJ units of length, mass, energy, and time,
as follows: $\sigma=\sigma_{AA}$, $m=m_{A}$,
$\varepsilon=\varepsilon_{AA}$, and, correspondingly,
$\tau=\sigma\sqrt{m/\varepsilon}$.   The MD simulations were carried
out using the LAMMPS code~\cite{Lammps} with the integration time
step $\triangle t_{MD}=0.005\,\tau$~\cite{Allen87}.

\vskip 0.05in


The sample preparation procedure is similar to the one employed in
the previous study on elastostatic loading along a single
direction~\cite{PriezELAST19}. Specifically, the binary mixture was
first equilibrated in a periodic box at the temperature
$T_{LJ}=1.0\,\varepsilon/k_B$ and zero pressure. Here, $k_B$ denotes
the Boltzmann constant. In all simulations, the temperature was
regulated via the Nos\'{e}-Hoover thermostat~\cite{Allen87,Lammps}.
After equilibration, the system was linearly cooled with the
computationally slow rate $10^{-5}\varepsilon/k_{B}\tau$ at zero
pressure to the reference temperature of $0.01\,\varepsilon/k_B$. In
the previous study, the glass transition temperature of the KA model
was found to be about $0.40\,\varepsilon/k_B$, when the mixture was
cooled with the rate $10^{-5}\varepsilon/k_{B}\tau$ at zero
pressure~\cite{Priez19one}.

\vskip 0.05in

\begin{figure}[t]
\includegraphics[width=12.0cm,angle=0]{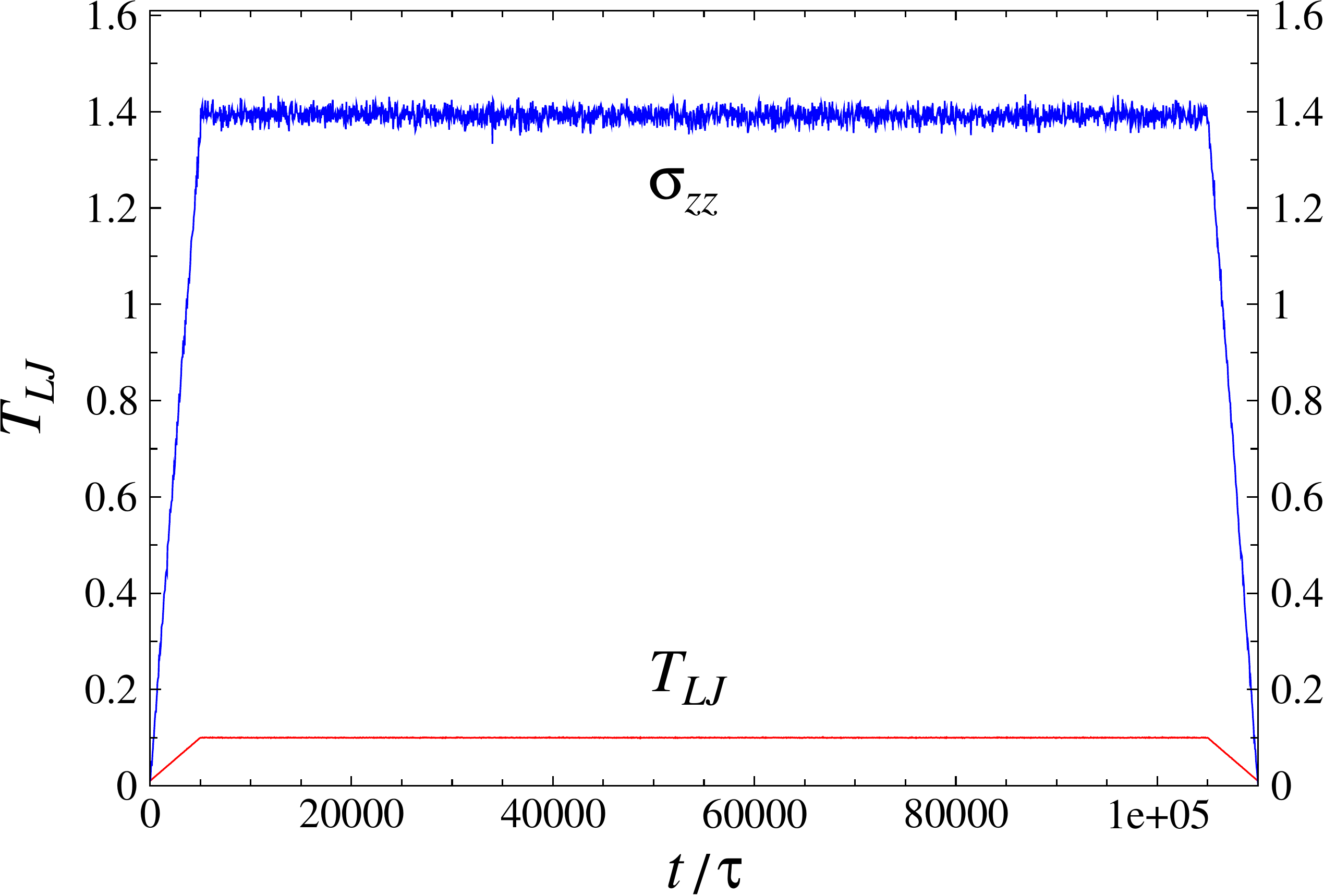}
\caption{(Color online) The temperature (left axis, red line) and
normal stress (right axis, blue line) profiles during the loading
protocol. The temperature varies linearly from
$T_{LJ}=0.01\,\varepsilon/k_B$ to $T_{a}=0.1\,\varepsilon/k_B$
during $5000\,\tau$, remains $T_{a}=0.1\,\varepsilon/k_B$ during the
annealing time $t_{a}=10^5\tau$, and then reduces to
$T_{LJ}=0.01\,\varepsilon/k_B$. The variation of the normal stress
is similar except that $\sigma_{zz}(0)=0$ and
$\sigma_{zz}(1.1\times10^5\tau)=0$. }
\label{fig:temper_stress_profs}
\end{figure}

The loading procedure includes heating the glass to the annealing
temperature $T_a$ and at the same time increasing the normal stress
during $5000\,\tau$, then annealing the system during the time
interval $t_a$ at a constant value of the normal stress, followed by
subsequent relocation of the glass to the reference state
($T_{LJ}=0.01\,\varepsilon/k_B$ and $P=0$) during $5000\,\tau$
(\textit{e.g.}, see Fig.\,\ref{fig:temper_stress_profs}). It was
previously shown that rejuvenation is enhanced during elastostatic
loading along a single direction at the annealing temperature
$T_{a}=0.1\,\varepsilon/k_B$~\cite{PriezELAST19}, and, therefore, in
the present study all simulations were carried out at this value of
$T_{a}$. Furthermore, the static stress was applied either along the
same direction or alternated along two and three spatial directions.
In all cases, the normal stress components perpendicular to the
imposed static stress were set to zero. During production runs, the
stress components, potential energy, atomic configurations, and
system dimensions were stored at $T_{LJ}=0.01\,\varepsilon/k_B$ and
$P=0$. The data were accumulated only for one realization of
disorder due to computational limitations.  For example, a typical
run during $t_a=2.4\times10^6\,\tau$ using $28$ processors required
about $360$ hours at the Ohio Supercomputer Cluster.

\section{Results}
\label{sec:Results}


It has long been realized that the rate at which a multicomponent
mixture is cooled across the glass transition point is an important
factor that determines the atomic structure and the potential energy
state of a glass~\cite{Greer16}. Hence, a more slowly annealed glass
is settled at a deeper energy level, whereas rapidly quenched
glasses typically have higher potential energy~\cite{Stillinger00}.
Moreover, depending on the amplitude of periodic strain deformation,
the glass can be either rejuvenated or relaxed~\cite{Lacks04}. In
the case of elastostatic loading, the glass can be rejuvenated via
collective rearrangements of atoms, predominantly near soft spots,
if the imposed stress is sufficiently large and the annealing
temperature is well below the glass transition
temperature~\cite{PriezELAST19}. Interestingly, a transition from
relaxed to rejuvenated states upon increasing static stress at about
half the glass transition temperature was reported in the recent MD
study~\cite{PriezELAST19}. In principle, the potential energy can be
further increased during elastostatic loading if the orientation of
the static stress is occasionally changed, thus allowing different
clusters of atoms to rearrange irreversibly. In the present study,
we explore the effect of rejuvenation in a well annealed binary
glass for three loading protocols; namely, when the static stress is
applied along a single direction or changed along two or three
spatial directions.

\vskip 0.05in


The results of the previous MD study on relaxation and rejuvenation
of binary glasses subjected to elastostatic
loading~\cite{PriezELAST19} provide guidance on the choice of
parameters for the loading protocol with alternating orientation of
the static stress. Thus, it was shown that rejuvenation is enhanced
if the annealing temperature is sufficiently below the glass
transition temperature $T_g\approx0.40\,\varepsilon/k_B$ at
$P=0$~\cite{PriezELAST19}. In the present study, all simulation
results are reported for the annealing temperature
$T_{LJ}=0.1\,\varepsilon/k_B$, at which the aging effects are
negligible during $t_a=2.4\times10^6\,\tau$~\cite{PriezELAST19}.
Moreover, it was also found that during compression of a well
annealed glass at the constant strain rate
$\dot{\varepsilon}=10^{-5}\,\tau^{-1}$ at
$T_{LJ}=0.1\,\varepsilon/k_B$, the peak value of the stress
overshoot is
$\sigma_Y\approx1.9\,\varepsilon\sigma^{-3}$~\cite{PriezELAST19}.
Therefore, the loading protocol in the present study includes four
values of the static stress below $\sigma_Y$; namely,
$0.5\,\varepsilon\sigma^{-3}$, $1.1\,\varepsilon\sigma^{-3}$,
$1.3\,\varepsilon\sigma^{-3}$, and $1.4\,\varepsilon\sigma^{-3}$.
Furthermore, the static stress was applied either along a single
($\hat{z}$) direction or alternated along two ($\hat{z}$ and
$\hat{x}$) and three ($\hat{z}$, $\hat{x}$, and $\hat{y}$)
directions. In the last two cases, the orientation of the static
stress was changed every $10^5\,\tau$.

\vskip 0.05in

\begin{figure}[t]
\includegraphics[width=12.0cm,angle=0]{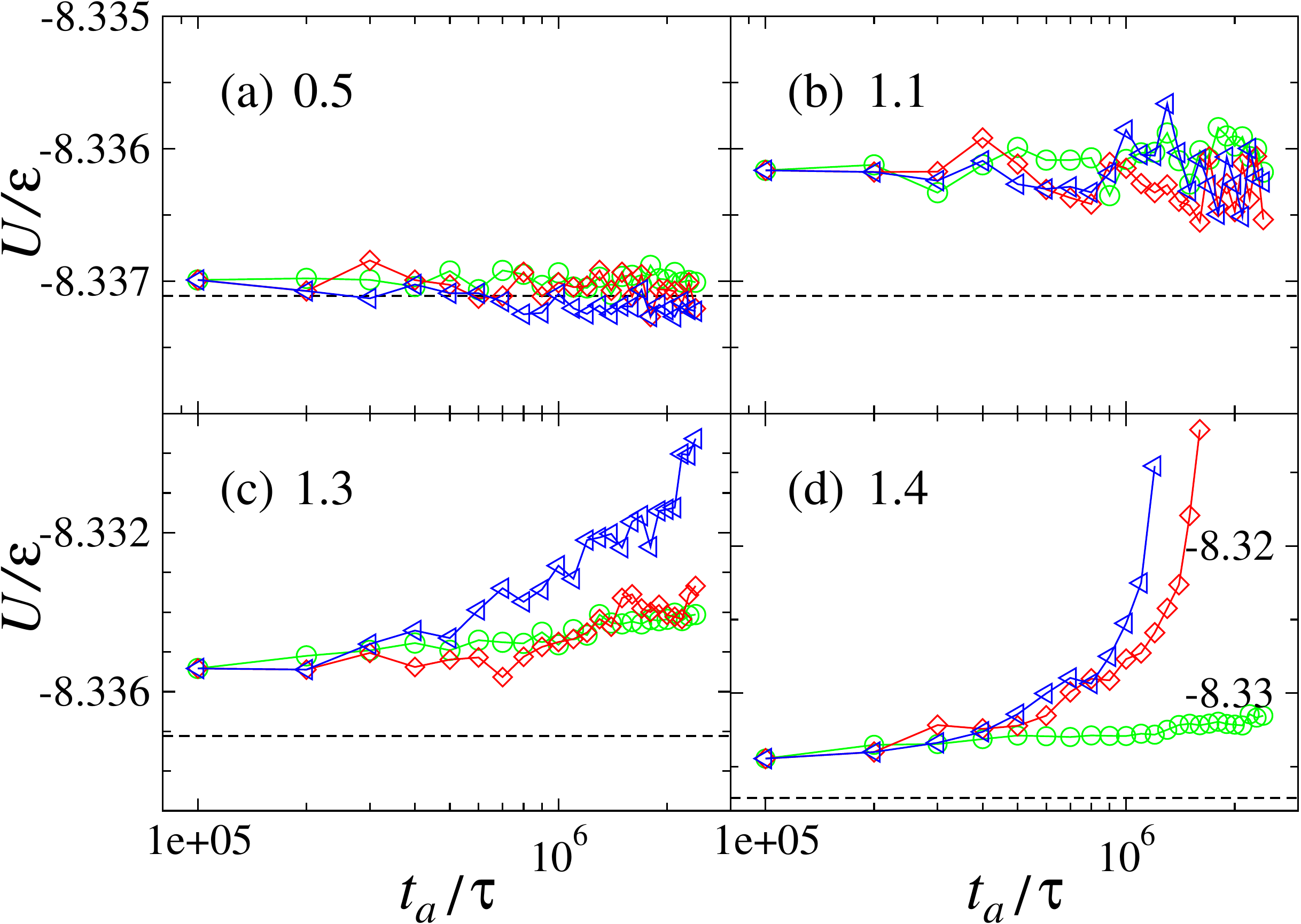}
\caption{(Color online) The potential energy per atom as a function
of the annealing time, $t_{a}/\tau$, when the static stress is (a)
$0.5\,\varepsilon\sigma^{-3}$, (b) $1.1\,\varepsilon\sigma^{-3}$,
(c) $1.3\,\varepsilon\sigma^{-3}$, and (d)
$1.4\,\varepsilon\sigma^{-3}$.  The static stress is applied along
the $\hat{z}$ direction (green circles), alternated along the
$\hat{z}$ and $\hat{x}$ directions (red diamonds), and alternated
along the $\hat{z}$, $\hat{x}$, and $\hat{y}$ directions (blue
triangles). The horizontal dashed lines indicate the potential
energy level $U\approx-8.337\,\varepsilon$ before elastostatic
loading was applied. }
\label{fig:poten_4_Pz}
\end{figure}

The dependence of the potential energy on the annealing time and
loading protocol is shown in Fig.\,\ref{fig:poten_4_Pz} for the
indicated values of the static stress.  Note that the data in
Fig.\,\ref{fig:poten_4_Pz} are taken at the reference state,
$T_{LJ}=0.01\,\varepsilon/k_B$ and $P=0$, whereas the static stress
was applied at the temperature $T_{a}=0.1\,\varepsilon/k_B$. As
shown in Fig.\,\ref{fig:poten_4_Pz}, the glass becomes rejuvenated
when the imposed static stress is sufficiently large (greater than
$0.5\,\varepsilon\sigma^{-3}$) and $t_a \lesssim
2.4\times10^6\,\tau$. Note that the increase in potential energy
with respect to the dashed line is larger for higher values of the
static stress.  This is consistent with the results of the previous
MD study on elastostatic loading where the static stress was applied
along a single direction~\cite{PriezELAST19}.  It can be clearly
seen in Fig.\,\ref{fig:poten_4_Pz}\,(c,\,d) that in the case of
alternating orientation of the static stress, the effect of
rejuvenation is significantly amplified.  This trend can be
rationalized by realizing that upon changing the orientation of the
static stress, the stress in the system becomes redistributed and a
different set of localized rearrangements can be facilitated (during
the time interval $10^5\,\tau$), leading to faster rejuvenation. As
shown in Fig.\,\ref{fig:poten_4_Pz}\,(c,\,d), the largest increase
in energy is attained when the static stress is alternated along all
three spatial directions.

\vskip 0.05in

\begin{figure}[t]
\includegraphics[width=12.0cm,angle=0]{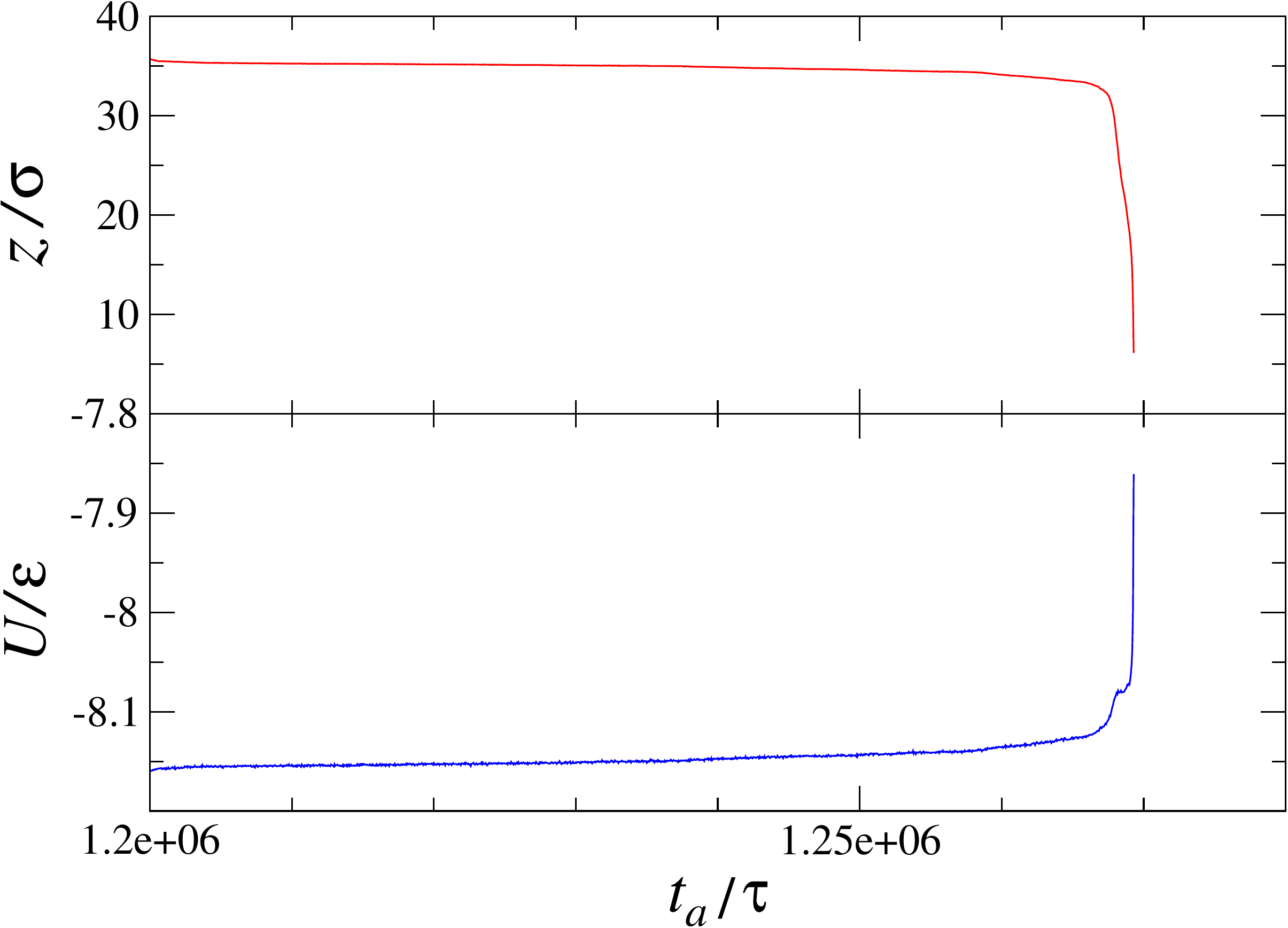}
\caption{(Color online) The time dependence of the system size along
the $\hat{z}$ axis (the upper panel) and the potential energy (the
lower panel) when the static stress is $1.4\,\varepsilon\sigma^{-3}$
and temperature is $T_{a}=0.1\,\varepsilon/k_B$. The loading
protocol consists of alternating static stress along all three
spatial directions (see text for details).}
\label{fig:collapse_Pzxy14}
\end{figure}

We further comment that during elastostatic loading at
$1.4\,\varepsilon\sigma^{-3}$ along alternating directions, the
samples were extensively compressed along the direction of the
applied stress. Therefore, the data are not reported for
$t_a\gtrsim1.2\times10^6\,\tau$ in Fig.\,\ref{fig:poten_4_Pz}\,(d).
For example, the time dependence of the potential energy and the
system size for alternating loading at $1.4\,\varepsilon\sigma^{-3}$
is presented in Fig.\,\ref{fig:collapse_Pzxy14}.  It can be seen in
Fig.\,\ref{fig:collapse_Pzxy14}\,(a) that, following a gradual
compression for $t_a\lesssim1.26\times10^6\,\tau$, the system
becomes significantly deformed during a relatively short time
interval $\approx\!2\times10^3\,\tau$. This plastic strain is
unrecoverable, and the corresponding change in the potential energy
is shown in Fig.\,\ref{fig:collapse_Pzxy14}\,(b).  In general, we
find that the loading time until plastic flow is reduced for larger
values of the static stress and loading protocols with alternating
stress orientation.

\vskip 0.05in


The spatial organization of stress-induced, irreversible
rearrangements of atoms can be analyzed via their nonaffine
displacements~\cite{Falk98}. More specifically, the nonaffine
measure for a particular atom can be computed by using the
transformation matrix $\mathbf{J}_i$, which linearly transforms
positions of neighboring atoms, and at the same time minimizes the
following expression:
\begin{equation}
D^2(t, \Delta t)=\frac{1}{N_i}\sum_{j=1}^{N_i}\Big\{
\mathbf{r}_{j}(t+\Delta t)-\mathbf{r}_{i}(t+\Delta t)-\mathbf{J}_i
\big[ \mathbf{r}_{j}(t) - \mathbf{r}_{i}(t)    \big] \Big\}^2,
\label{Eq:D2min}
\end{equation}
where $\Delta t$ is the time interval between two successive
configurations and the sum is taken over the neighboring atoms
within $1.5\,\sigma$ from the position $\mathbf{r}_{i}(t)$. It was
originally demonstrated by Falk and Langer that the nonaffine
measure is an excellent indicator of irreversible shear
transformations in deformed amorphous solids~\cite{Falk98}. More
recently, the appearance of clusters of atoms with relatively large
nonaffine displacements was connected to rejuvenation of
elastostatically loaded binary glasses~\cite{PriezELAST19}. In
recent years, the collective rearrangements of atoms in disordered
solids were also reported during
steady~\cite{Ma12,ShangWang14,Priez20tfic,Priez20star,Ozawa20} and
periodic~\cite{Priezjev16,Priezjev16a,Kawasaki16,Sastry17,Priezjev17,
Priezjev18,Priezjev18a,NVP18strload,NVP19alt,Priez19ba,Jana20,Priez20altY,Priez20del}
deformation and thermal
cycling~\cite{Priez18tcyc,Priez19T2000,Priez19T5000}.

\vskip 0.05in

\begin{figure}[t]
\includegraphics[width=12.cm,angle=0]{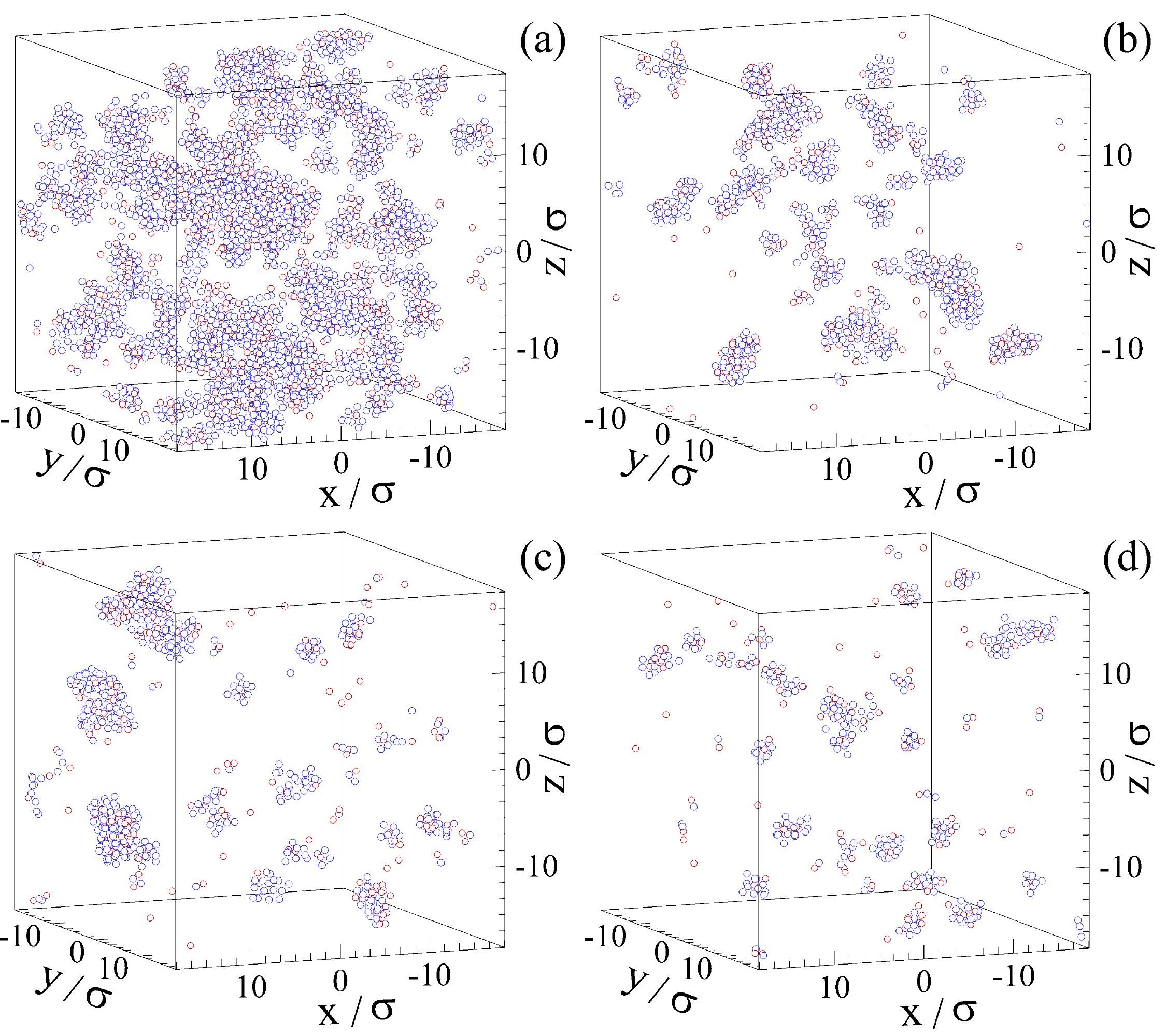}
\caption{(Color online) The snapshots of atomic positions with the
nonaffine measure $D^2(t, \Delta t)$: (a) $D^2(0,
6\times10^5\tau)>0.04\,\sigma^2$, (b) $D^2(6\times10^5\tau,
6\times10^5\tau)>0.04\,\sigma^2$, (c) $D^2(1.2\times10^6\tau,
6\times10^5\tau)>0.04\,\sigma^2$, and (d) $D^2(1.8\times10^6\tau,
6\times10^5\tau)>0.04\,\sigma^2$. The static stress
$1.3\,\varepsilon\sigma^{-3}$ is applied along a single ($\hat{z}$)
direction. The potential energy for the same sample is indicated by
green circles in Fig.\,\ref{fig:poten_4_Pz}\,(c).  }
\label{fig:snapshot_clusters_T01_Pz13}
\end{figure}
\begin{figure}[t]
\includegraphics[width=12.cm,angle=0]{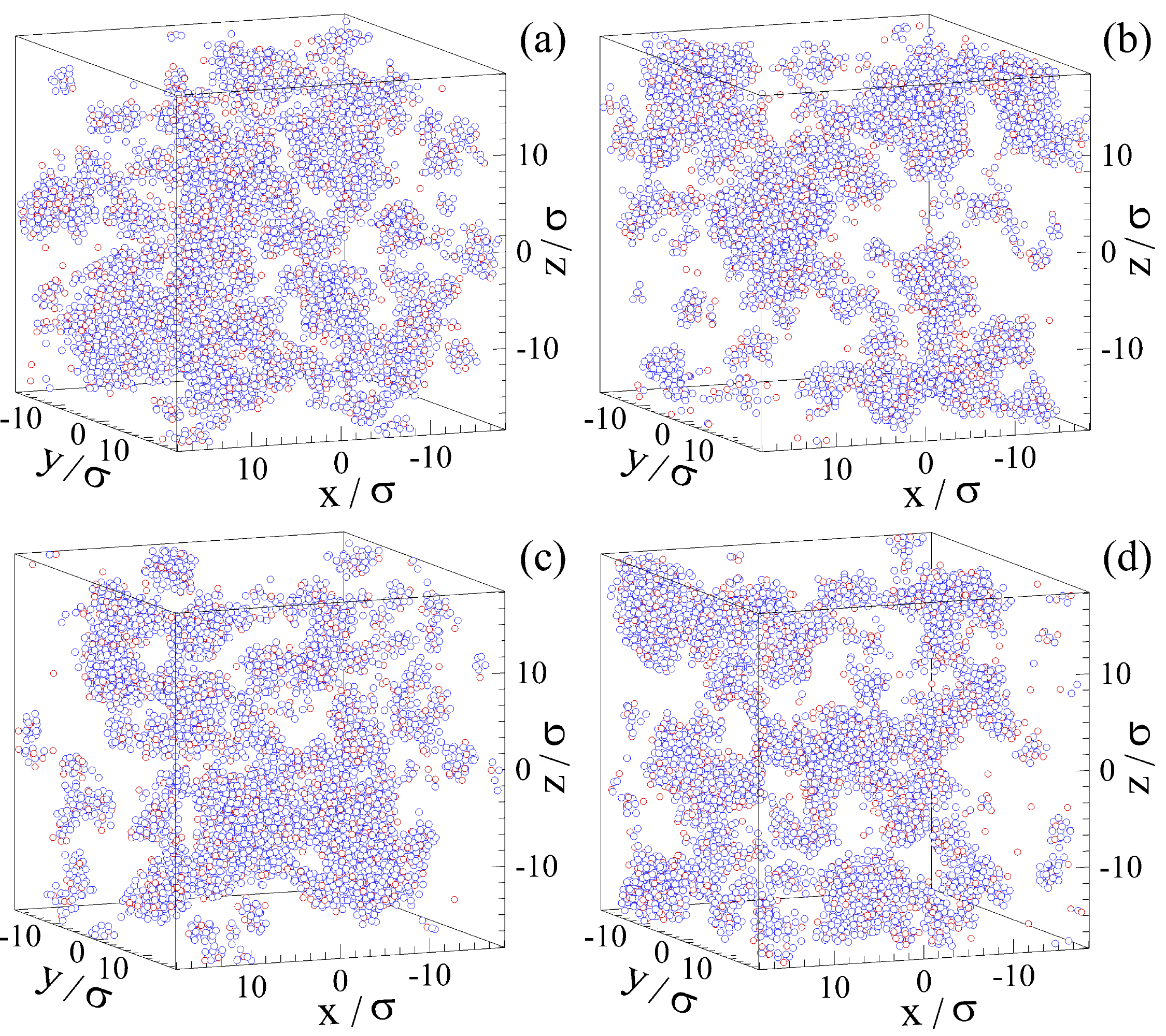}
\caption{(Color online) The atomic configurations with the nonaffine
measure $D^2(t, \Delta t)$: (a) $D^2(0,
6\times10^5\tau)>0.04\,\sigma^2$, (b) $D^2(6\times10^5\tau,
6\times10^5\tau)>0.04\,\sigma^2$, (c) $D^2(1.2\times10^6\tau,
6\times10^5\tau)>0.04\,\sigma^2$, and (d) $D^2(1.8\times10^6\tau,
6\times10^5\tau)>0.04\,\sigma^2$. The static stress
$1.3\,\varepsilon\sigma^{-3}$ is alternated along the $\hat{z}$ and
$\hat{x}$ directions. The same data are denoted by red diamonds in
Fig.\,\ref{fig:poten_4_Pz}\,(c).  }
\label{fig:snapshot_clusters_T01_Pzx13}
\end{figure}
\begin{figure}[t]
\includegraphics[width=12.cm,angle=0]{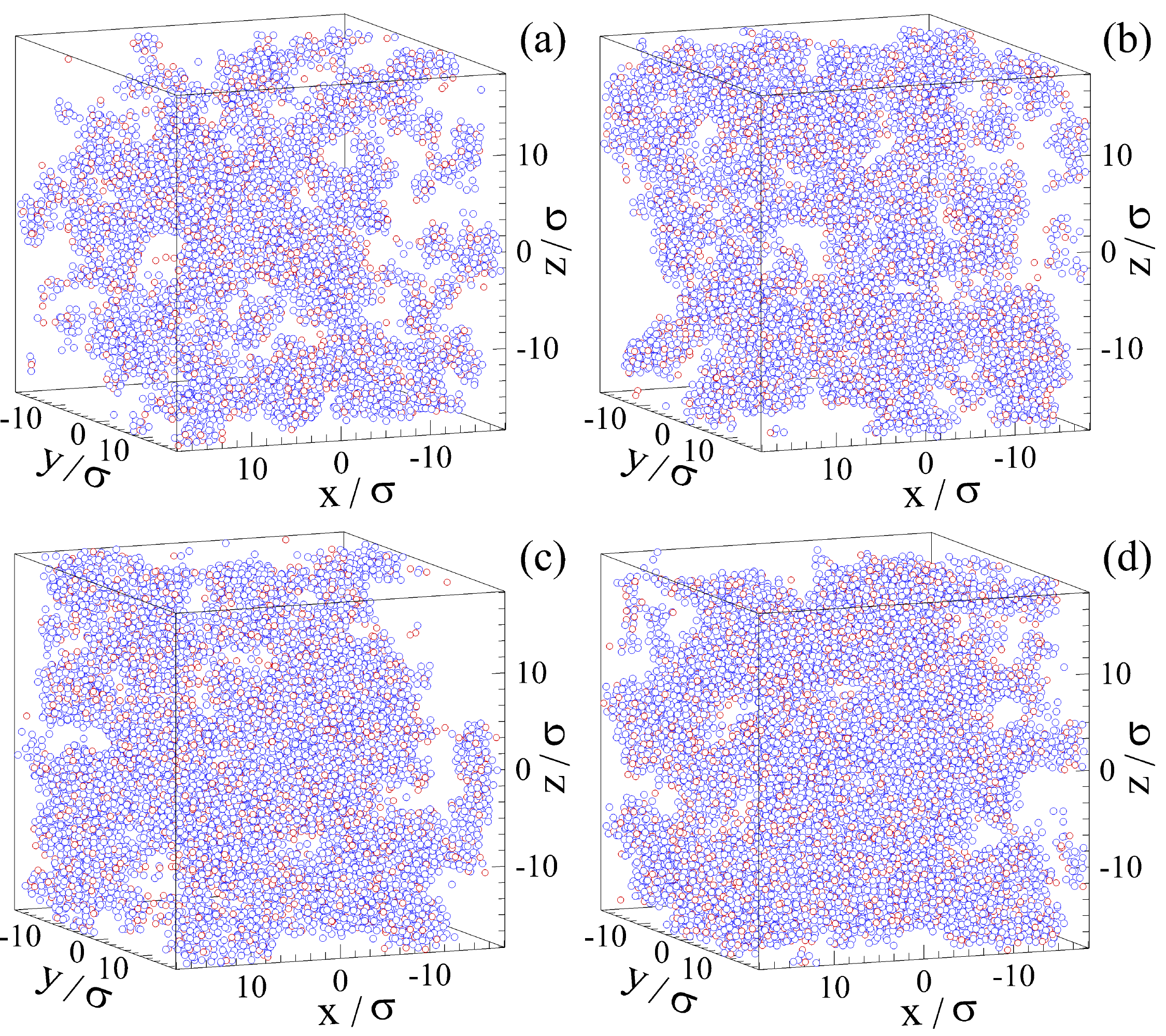}
\caption{(Color online) The positions of atoms with the nonaffine
measure $D^2(t, \Delta t)$: (a) $D^2(0,
6\times10^5\tau)>0.04\,\sigma^2$, (b) $D^2(6\times10^5\tau,
6\times10^5\tau)>0.04\,\sigma^2$, (c) $D^2(1.2\times10^6\tau,
6\times10^5\tau)>0.04\,\sigma^2$, and (d) $D^2(1.8\times10^6\tau,
6\times10^5\tau)>0.04\,\sigma^2$. The static stress
$1.3\,\varepsilon\sigma^{-3}$ is alternated along the $\hat{z}$,
$\hat{x}$, and $\hat{y}$ directions. The data are marked by blue
triangles in Fig.\,\ref{fig:poten_4_Pz}\,(c).  }
\label{fig:snapshot_clusters_T01_Pzxy13}
\end{figure}

The effects of loading time and stress orientation on plastic
rearrangements of atoms are presented in
Figures\,\ref{fig:snapshot_clusters_T01_Pz13}--\ref{fig:snapshot_clusters_T01_Pzxy13}.
In all cases, the value of the static stress is fixed to
$1.3\,\varepsilon\sigma^{-3}$, and the lag time in
Eq.\,(\ref{Eq:D2min}) is set $\Delta t=6\times10^5\tau$. The
reference times $t=0$, $6\times10^5\tau$, $1.2\times10^6\tau$, and
$1.8\times10^6\tau$ are chosen within the rejuvenation regime,
$0\leqslant t_a \leqslant 2.4\times10^6\,\tau$, as shown in
Fig.\,\ref{fig:poten_4_Pz}\,(c). The loading protocol consists of
the static stress applied either along a single ($\hat{z}$)
direction (see Fig.\,\ref{fig:snapshot_clusters_T01_Pz13}) or
alternated along two ($\hat{z}$ and $\hat{x}$) directions (see
Fig.\,\ref{fig:snapshot_clusters_T01_Pzx13}) or all three directions
(see Fig.\,\ref{fig:snapshot_clusters_T01_Pzxy13}). It is clearly
seen in
Figs.\,\ref{fig:snapshot_clusters_T01_Pz13}--\ref{fig:snapshot_clusters_T01_Pzxy13}
that the size of clusters of atoms with large nonaffine
displacements increases when an additional spatial dimension for the
stress orientation is included in the loading protocol. Note also
that upon loading along a single direction, a large fraction of the
system deforms plastically during the first $\Delta
t=6\times10^5\tau$, but only a few isolated clusters are present at
longer times, as shown in
Fig.\,\ref{fig:snapshot_clusters_T01_Pz13}. This observation
correlates with a steep increase and subsequent slow growth of the
potential energy, as indicated by green circles in
Fig.\,\ref{fig:poten_4_Pz}\,(c).  By contrast, when the stress
orientation is alternated along all three spatial directions, most
of the atoms undergo large nonaffine displacements (larger than the
typical cage size $\approx\!0.1\,\sigma$), and the number of
rearrangements increases with time (see
Fig.\,\ref{fig:snapshot_clusters_T01_Pzxy13}), which corresponds to
the enhanced rejuvenation shown by blue triangles in
Fig.\,\ref{fig:poten_4_Pz}\,(c).

\vskip 0.05in

\begin{figure}[t]
\includegraphics[width=12.cm,angle=0]{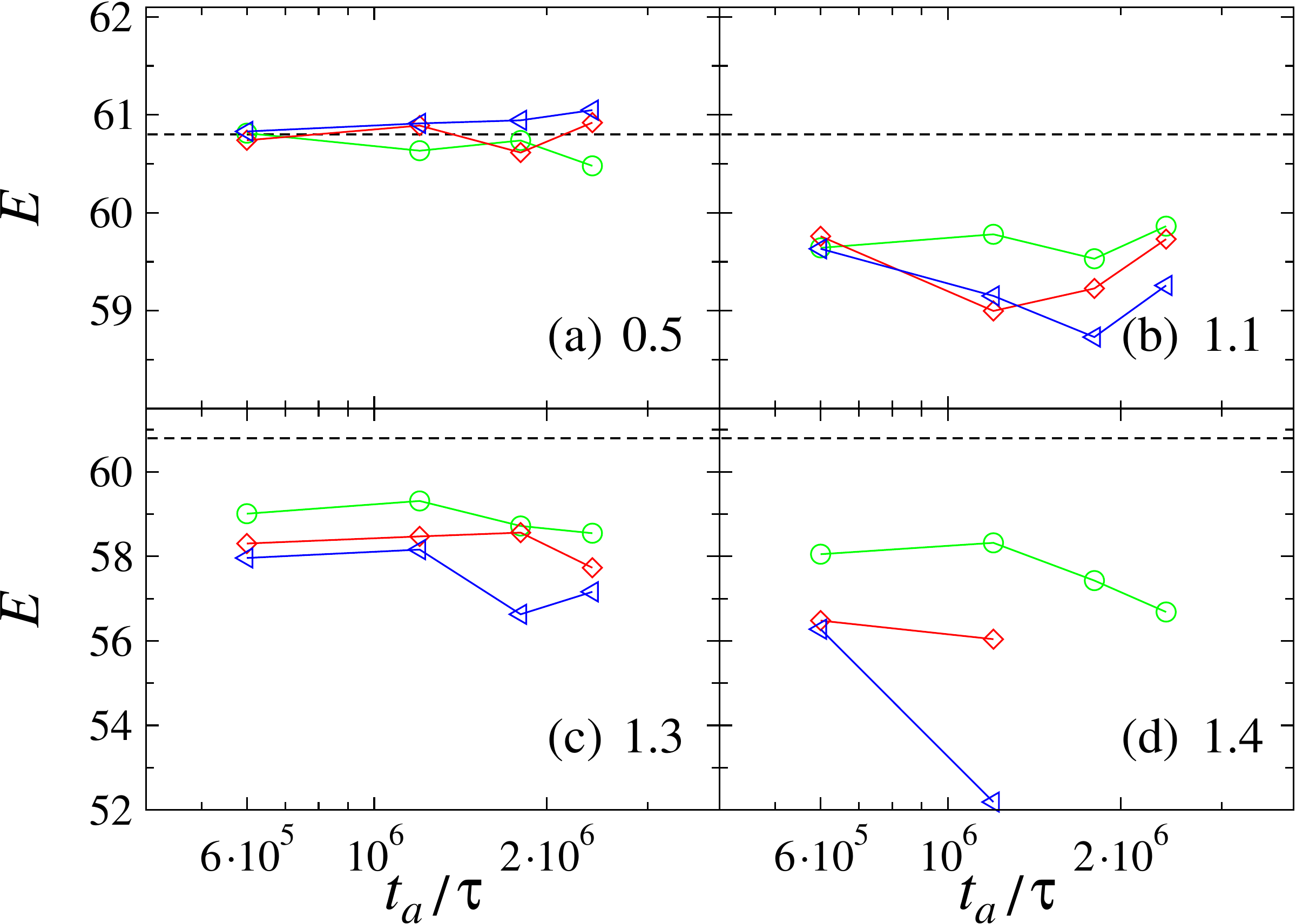}
\caption{(Color online) The variation of the elastic modulus $E$ (in
units of $\varepsilon\sigma^{-3}$) as a function of the annealing
time, $t_{a}/\tau$, for the values of the static stress (a)
$0.5\,\varepsilon\sigma^{-3}$, (b) $1.1\,\varepsilon\sigma^{-3}$,
(c) $1.3\,\varepsilon\sigma^{-3}$, and (d)
$1.4\,\varepsilon\sigma^{-3}$. The elastostalic loading is applied
along the $\hat{z}$ direction (green circles), alternated along the
$\hat{z}$ and $\hat{x}$ directions (red diamonds), and alternated
along the $\hat{z}$, $\hat{x}$, and $\hat{y}$ directions (blue
triangles). The dashed black lines denote the elastic modulus
$E=60.80\,\varepsilon\sigma^{-3}$ at $T_{LJ}=0.01\,\varepsilon/k_B$
before elastostalic loading.  Note that the vertical scales are
different in the upper and lower panels. }
\label{fig:E}
\end{figure}
\begin{figure}[t]
\includegraphics[width=12.cm,angle=0]{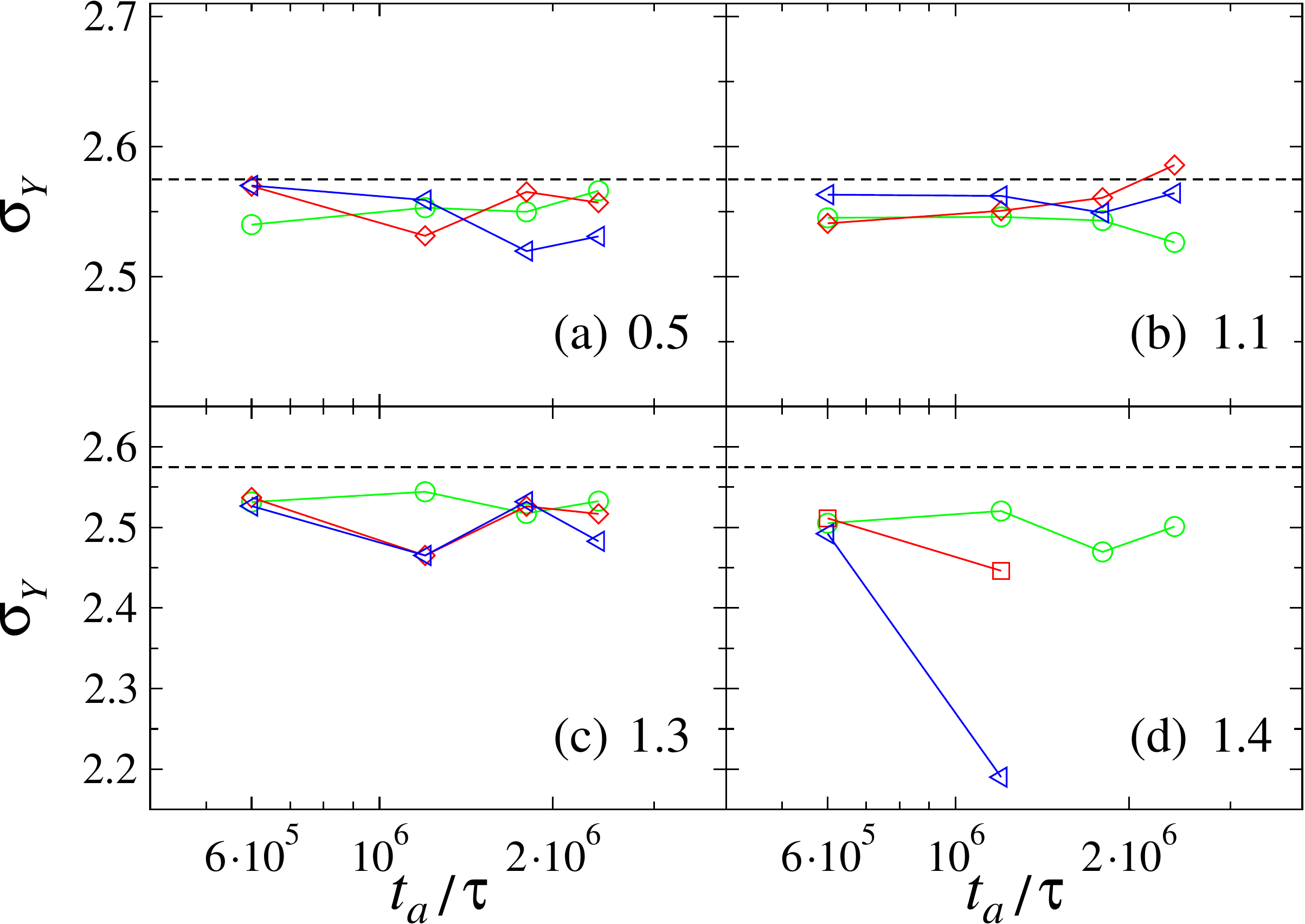}
\caption{(Color online) The yielding peak $\sigma_Y$ (in units of
$\varepsilon\sigma^{-3}$) versus the annealing time for the static
stress (a) $0.5\,\varepsilon\sigma^{-3}$, (b)
$1.1\,\varepsilon\sigma^{-3}$, (c) $1.3\,\varepsilon\sigma^{-3}$,
and (d) $1.4\,\varepsilon\sigma^{-3}$. The static stress is imposed
along the $\hat{z}$ direction (green circles), alternated along the
$\hat{z}$ and $\hat{x}$ directions (red diamonds), and alternated
along the $\hat{z}$, $\hat{x}$, and $\hat{y}$ directions (blue
triangles).  The dashed lines indicate the peak value of the stress
overshoot $\sigma_Y=2.57\,\varepsilon\sigma^{-3}$ before static
loading.  }
\label{fig:sigY}
\end{figure}

We next present the elastic modulus, $E$, in Fig.\,\ref{fig:E} and
the yield stress, $\sigma_Y$, in Fig.\,\ref{fig:sigY} for the subset
of data points shown in Fig.\,\ref{fig:poten_4_Pz}. The mechanical
properties reported in Figures\,\ref{fig:E} and \ref{fig:sigY} were
computed during startup continuous compression with the constant
strain rate $\dot{\varepsilon}=10^{-5}\,\tau^{-1}$.  More
specifically, the elastic modulus was calculated from the best
linear fit to the data at small strain, $\varepsilon\leqslant0.01$,
whereas the yield stress is defined by the peak value of the stress
overshoot at about 5\% strain.  The samples were strained along the
$\hat{x}$, $\hat{y}$, and $\hat{z}$ directions, and then the data
for $E$ and $\sigma_Y$ were averaged over three spatial directions.
The horizontal dashed lines in Figs.\,\ref{fig:E} and \ref{fig:sigY}
indicate the values of $E$ and $\sigma_Y$ before elastostatic
loading was applied.  As expected, the general trend for the time
dependence of $E$ and $\sigma_Y$ is inversely correlated with the
variation of the potential energy shown in
Fig.\,\ref{fig:poten_4_Pz}. In particular, it can be observed that
the yield stress is only slightly reduced for the static loading at
$0.5\,\varepsilon\sigma^{-3}$ and $1.1\,\varepsilon\sigma^{-3}$,
while the effect of alternating loading becomes significant at
larger values of the static stress, $1.3\,\varepsilon\sigma^{-3}$
and $1.4\,\varepsilon\sigma^{-3}$ (see Fig.\,\ref{fig:sigY}). Thus,
the maximum decrease of the yield stress due to elastostatic loading
at $1.4\,\varepsilon\sigma^{-3}$ along three directions is about
15\% at $t=1.2\times10^6\tau$, as shown in
Fig.\,\ref{fig:sigY}\,(c).  Altogether, these results demonstrate
that both elastic modulus and yield stress become reduced upon
loading at sufficiently large values of the static stress; and the
effect is amplified if the loading orientation is periodically
alternated along two or three spatial directions.

\section{Conclusions}

In summary, we investigated the effects of elastostatic loading and
alternating orientation of the static stress on the potential energy
states and mechanical properties of amorphous alloys using molecular
dynamics simulations.  The well annealed alloy was prepared by
slowly cooling a binary mixture at zero pressure from the liquid
state to a temperature well below the glass transition temperature.
After annealing, the binary glass was subjected to prolonged
elastostatic compression either along a single direction or along
two and three alternating directions. It was demonstrated that
elastostatic loading at sufficiently large values of stress induces
collective plastic events and significant rejuvenation.   Moreover,
upon introducing an alternating stress orientation in the loading
protocol, the potential energy is further increased with respect to
the untreated sample, and the typical size of plastic rearrangements
becomes comparable with the system size.    Finally, the mechanical
properties were probed by continuously compressing the treated
samples at a constant strain rate.   We found that both the elastic
modulus and the peak value of the stress overshoot are reduced with
increasing loading time or alternating stress orientation.

\section*{Acknowledgments}

Financial support from the National Science Foundation (CNS-1531923)
and the ACS Petroleum Research Fund (60092-ND9) is gratefully
acknowledged.  The article was prepared within the framework of the
HSE University Basic Research Program and funded in part by the
Russian Academic Excellence Project `5-100'. The molecular dynamics
simulations were performed using the LAMMPS open-source code
developed at Sandia National Laboratories~\cite{Lammps}. The
numerical simulations were carried out at the Wright State
University's Computing Facility and the Ohio Supercomputer Center.



%

%

%

%

%

%

%

%

\bibliographystyle{prsty}

\begin{thebibliography}{99}


\bibitem{Forming20}    Z. Li, Z. Huang, F. Sun, X. Li, and J. Ma,
                       Forming of metallic glasses: mechanisms and processes,
                       Materials Today Advances {\bf 7}, 100077 (2020).

\bibitem{Greer16}      Y. Sun, A. Concustell, and A.~L. Greer,
                       Thermomechanical processing of metallic glasses: extending the range of the glassy state,
                       Nat. Rev. Mater. {\bf 1}, 16039 (2016).

\bibitem{Greer20hard}  J. Pan, Yu.~P. Ivanov, W.~H. Zhou, Y. Li, and A.~L. Greer,
                       Strain-hardening and suppression of shear-banding in rejuvenated bulk metallic glass,
                       Nature {\bf 578}, 559 (2020).




\bibitem{Ogata15}      M. Wakeda, J. Saida, J. Li, and S. Ogata,
                       Controlled rejuvenation of amorphous metals with thermal processing,
                       Sci. Rep. {\bf 5}, 10545 (2015).

\bibitem{Maass18}      S. Kuchemann, P.~M. Derlet, C. Liu, D. Rosenthal, G. Sparks, W.~S. Larson, and R. Maass,
                       Energy storage in metallic glasses via flash annealing,
                       Adv. Funct. Mater. {\bf 28}, 1805385 (2018).

\bibitem{Priez19one}   N.~V. Priezjev,
                       Atomistic modeling of heat treatment processes for tuning the mechanical properties of disordered solids,
                       J. Non-Cryst. Solids {\bf 518}, 128 (2019).

\bibitem{WangTang20}   M. Wang, H. Liu, J. Li, Q. Jiang, W. Yang, and C. Tang,
                       Thermal-pressure treatment for tuning the atomic structure of metallic glass Cu-Zr,
                       J. Non-Cryst. Solids {\bf 535}, 119963 (2020).





\bibitem{Park08}       K.-W. Park, C.-M. Lee, M. Wakeda, Y. Shibutani, M.~L. Falk, and J.-C. Lee,
                       Elastostatically induced structural disordering in amorphous alloys,
                       Acta Materialia {\bf 56}, 5440 (2008).

\bibitem{Jae-Chul09}   C.-M. Lee, K.-W. Park, B.-J. Lee, Y. Shibutani and J.-C. Lee,
                       Structural disordering of amorphous alloys: A molecular dynamics analysis,
                       Scripta Materialia {\bf 61}, 911 (2009).

\bibitem{Tong13}       Y. Tong, W. Dmowski, Y. Yokoyama, G. Wang, P.~K. Liaw, and T. Egami,
                       Recovering compressive plasticity of bulk metallic glasses by high-temperature creep,
                       Scripta Materialia {\bf 69} 570 (2013).


\bibitem{Wang15}       Y.~M. Wang, M. Zhang and L. Liu,
                       Mechanical annealing in the homogeneous deformation of bulk
                       metallic glass under elastostatic compression,
                       Scripta Materialia {\bf 102}, 67 (2015).

\bibitem{Bai15}        L.~Z. Zhao, R.~J. Xue, Y.~Z. Li, W.~H. Wang, and H.~Y. Bai,
                       Revealing localized plastic flow in apparent elastic region before yielding in metallic glasses,
                       J. Appl. Phys. {\bf 118}, 244901 (2015).

\bibitem{GreerSun16}   A.~L. Greer, and Y.~H. Sun,
                       Stored energy in metallic glasses due to strains within the elastic limit,
                       Philos. Mag. {\bf 96}, 1643 (2016).

\bibitem{Zhang17}      M. Zhang, Y.~M. Wang, F.~X. Li, S.~Q. Jiang, M.~Z. Li and L. Liu,
                       Mechanical relaxation-to-rejuvenation transition in a Zr-based bulk metallic glass,
                       Sci. Rep. {\bf 7}, 625 (2017).

\bibitem{PanGreer18}   J. Pan, Y.~X. Wang, Q. Guo, D. Zhang, A.~L. Greer, and Y. Li,
                       Extreme rejuvenation and softening in a bulk metallic glass,
                       Nat. Commun. {\bf 9}, 560 (2018).

\bibitem{Samavatian19} M. Samavatian, R. Gholamipour, A.~A. Amadeh, and S. Mirdamadi,
                       Role of tensile elastostatic loading on atomic structure and mechanical
                       properties of Zr$_{55}$Cu$_{30}$Ni$_{5}$Al$_{10}$ bulk metallic glass,
                       Mater. Sci. Eng. A  {\bf 753}, 218 (2019).

\bibitem{PriezELAST19} N.~V. Priezjev,
                       Aging and rejuvenation during elastostatic loading of amorphous alloys:
                       A molecular dynamics simulation study,
                       Comput. Mater. Sci. {\bf 168}, 125 (2019).



\bibitem{Ketov15}      S.~V. Ketov, Y.~H. Sun, S. Nachum, Z. Lu, A. Checchi, A.~R. Beraldin, H.~Y. Bai,
                       W.~H. Wang, D.~V. Louzguine-Luzgin, M.~A. Carpenter, and A.~L. Greer,
                       Rejuvenation of metallic glasses by non-affine thermal strain,
                       Nature {\bf 524}, 200 (2015).

\bibitem{Guo19}        W. Guo, J. Saida, M. Zhao, S. Lu, and S. Wu,
                       Rejuvenation of Zr-based bulk metallic glass matrix composite upon deep cryogenic cycling,
                       Materials Letters {\bf 247}, 135 (2019).

\bibitem{Mirdamadi19}  M. Samavatian, R. Gholamipour, A.~A. Amadeh and S. Mirdamadi,
                       Correlation between plasticity and atomic structure evolution of a rejuvenated bulk metallic glass,
                       Metall. Mater. Trans. A {\bf 50}, 4743 (2019).



\bibitem{Lo10}         Y.~C. Lo, H.~S. Chou, Y.~T. Cheng, J.~C. Huang, J.~R. Morris, P.~K. Liaw,
                       Structural relaxation and self-repair behavior in nano-scaled Zr-Cu metallic glass
                       under cyclic loading: Molecular dynamics simulations,
                       Intermetallics {\bf 18}, 954 (2010).

\bibitem{Priezjev13}   N.~V. Priezjev,
                       Heterogeneous relaxation dynamics in amorphous materials under cyclic loading,
                       Phys. Rev. E {\bf 87}, 052302 (2013).

\bibitem{Sastry13}     D. Fiocco, G. Foffi, and S. Sastry,
                       Oscillatory athermal quasistatic deformation of a model glass,
                       Phys. Rev. E {\bf 88}, 020301(R) (2013).

\bibitem{Reichhardt13} I. Regev, T. Lookman, and C. Reichhardt,
                       Onset of irreversibility and chaos in amorphous solids under periodic shear,
                       Phys. Rev. E {\bf 88}, 062401 (2013).

\bibitem{Priezjev14}   N.~V. Priezjev,
                       Dynamical heterogeneity in periodically deformed polymer glasses,
                       Phys. Rev. E {\bf 89}, 012601 (2014).

\bibitem{IdoNature15}  I. Regev, J. Weber, C. Reichhardt, K.~A. Dahmen, and T. Lookman,
                       Reversibility and criticality in amorphous solids,
                       Nat. Commun. {\bf 6}, 8805 (2015).

\bibitem{Priezjev16}   N.~V. Priezjev,
                       Reversible plastic events during oscillatory deformation of amorphous solids,
                       Phys. Rev. E {\bf 93}, 013001 (2016).

\bibitem{Priezjev16a}  N.~V. Priezjev,
                       Nonaffine rearrangements of atoms in deformed and quiescent binary glasses,
                       Phys. Rev. E {\bf 94}, 023004 (2016).

\bibitem{Kawasaki16}   T. Kawasaki and L. Berthier,
                       Macroscopic yielding in jammed solids is accompanied by a non-equilibrium
                       first-order transition in particle trajectories,
                       Phys. Rev. E {\bf 94}, 022615 (2016).

\bibitem{Sastry17}     P. Leishangthem, A.~D.~S. Parmar, and S. Sastry,
                       The yielding transition in amorphous solids under oscillatory shear deformation,
                       Nat. Commun. {\bf 8}, 14653 (2017).

\bibitem{Priezjev17}   N.~V. Priezjev,
                       Collective nonaffine displacements in amorphous materials during large-amplitude oscillatory shear,
                       Phys. Rev. E {\bf 95}, 023002 (2017).

\bibitem{Priezjev18}   N.~V. Priezjev,
                       Molecular dynamics simulations of the mechanical annealing process in
                       metallic glasses: Effects of strain amplitude and temperature,
                       J. Non-Cryst. Solids {\bf 479}, 42 (2018).

\bibitem{Priezjev18a}  N.~V. Priezjev,
                       The yielding transition in periodically sheared binary glasses at finite temperature,
                       Comput. Mater. Sci. {\bf 150}, 162 (2018).

\bibitem{NVP18strload} N.~V. Priezjev,
                       Slow relaxation dynamics in binary glasses during stress-controlled,
                       tension-compression cyclic loading,
                       Comput. Mater. Sci. {\bf 153}, 235 (2018).

\bibitem{SastryPRX19}  A.~D.~S. Parmar, S. Kumar, and S. Sastry,
                       Strain localization above the yielding point in cyclically deformed glasses,
                       Phys. Rev. X {\bf 9}, 021018 (2019).

\bibitem{PriMakrho05}  N.~V. Priezjev and M.~A. Makeev,
                       The influence of periodic shear on structural relaxation and pore redistribution in binary glasses,
                       J. Non-Cryst. Solids {\bf 506}, 14 (2019).

\bibitem{PriMakrho09}  N.~V. Priezjev and M.~A. Makeev,
                       Structural transformations during periodic deformation of low-porosity amorphous materials,
                       Modelling Simul. Mater. Sci. Eng. {\bf 27}, 025004 (2019).

\bibitem{NVP19alt}     N.~V. Priezjev,
                       Accelerated relaxation in disordered solids under cyclic loading with alternating shear orientation,
                       J. Non-Cryst. Solids {\bf 525}, 119683 (2019).

\bibitem{Priez19ba}    N.~V. Priezjev,
                       Shear band formation in amorphous materials under oscillatory shear deformation,
                       Metals {\bf 10}, 300 (2020).

\bibitem{Jana20}       P.~K. Jana and N.~V. Priezjev,
                       Structural relaxation in amorphous materials under cyclic tension-compression loading,
                       J. Non-Cryst. Solids {\bf 540}, 120098 (2020).

\bibitem{Peng20}       H. Li, H. Liu, and H. Peng,
                       Atomic dynamics under oscillatory shear in metallic glasses,
                       J. Non-Cryst. Solids {\bf 539}, 120069 (2020).

\bibitem{Priez20altY}  N.~V. Priezjev,
                       Alternating shear orientation during cyclic loading facilitates yielding in amorphous materials
                       (2020). arXiv:2001.06853

\bibitem{Pelletier20}  G.-J. Lyu, J.-C. Qiao, Y. Yao, J.-M. Pelletier, D. Rodney, J. Morthomas, and C. Fusco,
                       Dynamic correspondence principle in the viscoelasticity of metallic glasses,
                       Scripta Materialia {\bf 174}, 39 (2020).

\bibitem{Priez20del}   N.~V. Priezjev,
                       A delayed yielding transition in mechanically annealed binary glasses at finite temperature,
                       J. Non-Cryst. Solids {\bf 548}, 120324 (2020).




\bibitem{KobAnd95}     W. Kob and H.~C. Andersen,
                       Testing mode-coupling theory for a supercooled binary Lennard-Jones mixture:
                       The van Hove correlation function,
                       Phys. Rev. E {\bf 51}, 4626 (1995).

\bibitem{Dyre19}       T.~S. Ingebrigtsen, J.~C. Dyre, T.~B. Schroder, and C.~P. Royall,
                       Crystallization instability in glass-forming mixtures,
                       Phys. Rev. X {\bf 9}, 031016 (2019).

\bibitem{Weber85}      T.~A. Weber and F.~H. Stillinger,
                       Local order and structural transitions in amorphous metal-metalloid alloys,
                       Phys. Rev. B {\bf 31}, 1954 (1985).


\bibitem{Lammps}       S.~J. Plimpton,
                       Fast parallel algorithms for short-range molecular dynamics,
                       J. Comp. Phys. {\bf 117}, 1 (1995).


\bibitem{Allen87}      M.~P. Allen and D.~J. Tildesley,
                       {\it Computer Simulation of Liquids} (Clarendon, Oxford, 1987).


\bibitem{Stillinger00} M. Utz, P.~G. Debenedetti, and F.~H. Stillinger,
                       Atomistic simulation of aging and rejuvenation in glasses,
                       Phys. Rev. Lett. {\bf 84}, 1471 (2000).

\bibitem{Lacks04}      D.~J. Lacks and M.~J. Osborne,
                       Energy landscape picture of overaging and rejuvenation in a sheared glass,
                       Phys. Rev. Lett. {\bf 93}, 255501 (2004).

\bibitem{Falk98}       M.~L. Falk and J.~S. Langer,
                       Dynamics of viscoplastic deformation in amorphous solids,
                       Phys. Rev. E {\bf 57}, 7192 (1998).



\bibitem{Ma12}         J. Ding, Y.~Q. Cheng, and E. Ma,
                       Correlating local structure with inhomogeneous elastic deformation in a metallic glass,
                       Appl. Phys. Lett. {\bf 101}, 121917 (2012).

\bibitem{ShangWang14}  B.~S. Shang, M.~Z. Li, Y.~G. Yao, Y.~J. Lu, and W.~H. Wang,
                       Evolution of atomic rearrangements in deformation in metallic glasses,
                       Phys. Rev. E {\bf 90}, 042303 (2014).

\bibitem{Priez20tfic}  N.~V. Priezjev,
                       The effect of thermal history on the atomic structure and mechanical properties of amorphous alloys,
                       Comput. Mater. Sci. {\bf 174}, 109477 (2020).

\bibitem{Priez20star}  N.~V. Priezjev,
                       Spatiotemporal analysis of nonaffine displacements in disordered solids sheared across the yielding point,
                       Metall. Mater. Trans. A {\bf 51}, 3713 (2020).

\bibitem{Ozawa20}      M. Singh, M. Ozawa, and L. Berthier,
                       Brittle yielding of amorphous solids at finite shear rates,
                       Phys. Rev. Mater. {\bf 4}, 025603 (2020).




\bibitem{Priez18tcyc}  N.~V. Priezjev,
                       The effect of cryogenic thermal cycling on aging, rejuvenation,
                       and mechanical properties of metallic glasses,
                       J. Non-Cryst. Solids {\bf 503}, 131 (2019).

\bibitem{Priez19T2000} Q.-L. Liu and N.~V. Priezjev,
                       The influence of complex thermal treatment on mechanical properties of amorphous materials,
                       Comput. Mater. Sci. {\bf 161}, 93 (2019).

\bibitem{Priez19T5000} N.~V. Priezjev,
                       Potential energy states and mechanical properties of thermally cycled binary glasses,
                       J. Mater. Res. {\bf 34}, 2664 (2019).



\end{thebibliography}

\end{document}